\documentclass[usenatbib]{mn2e}
\usepackage[pdftex]{graphicx}
\usepackage{amsmath}

\def\CN2{\mbox{$C_N^2 \ $}}
\def\CT2{\mbox{$C_T^2 \ $}}
\def\tauO{\mbox{$\tau_{0} \ $}}
\def\thetaO{\mbox{$\theta_{0} \ $}}

\title[MOSE: atmospherical parameters vertical stratification]{MOSE:  optical turbulence and atmospherical parameters operational forecast at ESO ground-based sites. \\
I: Overview and atmospherical parameters vertical stratification on [0-20] km.}
\author[E. Masciadri et al.]{E. Masciadri$^{1}$\thanks{E-mail:
     masciadri@arcetri.astro.it}, F. Lascaux$^{1}$, L. Fini$^{1}$\\ 
$^{1}$INAF Osservatorio Astrofisico di Arcetri, Largo Enrico Fermi 5, I-501 25 Florence, Italy}

\begin{document}
\newcommand{\cn}{$C_N^2$}
\label{firstpage}
\date{Accepted 2013 September 7, Received 2013 September 6; in original form 2013 July 25}  
\pagerange{\pageref{firstpage}--\pageref{lastpage}}
\pubyear{2013}
\maketitle
\begin{abstract}
We present the overview of the MOSE project (MOdeling ESO Sites) aiming at proving the feasibility of the forecast of the classical atmospherical parameters (wind speed intensity and direction, temperature, relative humidity) and the optical turbulence OT ($\CN2$ profiles and the most relevant integrated astro-climatic parameters derived from the $\CN2$: the seeing $\varepsilon$, 
the isoplanatic angle $\thetaO$, the wavefront coherence time $\tauO$) above the two ESO ground-based sites of Cerro Paranal 
and Cerro Armazones. The final outcome of the study is to investigate the opportunity to implement an automatic system for the 
forecast of these parameters at these sites. In this paper we present results related to the Meso-Nh model ability in reconstructing the vertical stratification of the atmospherical parameters along the 20 km above the ground. The very satisfactory performances shown by the model in reconstructing most of these parameters (and in particular the wind speed) put this tool of investigation as the most suitable to be used in astronomical observatories to support AO facilities and to calculate the temporal evolution of the wind speed and the wavefront coherence time at whatever temporal sampling. The further great advantage of this solution is that such estimates can be available in advance (order of some hours) with respect to the time of interest.

\end{abstract}
\begin{keywords} turbulence - atmospheric effects - methods: numerical - method: data analysis - balloons - site testing
\end{keywords}
\section{Introduction}

The MOSE project (MOdeling ESO Sites) aims at proving the feasibility of the forecast of all the most relevant classical atmospherical parameters 
for astronomical applications (wind speed intensity and direction, temperature, relative humidity) and the optical turbulence OT ($\CN2$ profiles) 
with the integrated astro-climatic parameters derived from the $\CN2$ i.e. the seeing ($\varepsilon$), 
the isoplanatic angle ($\thetaO$), the wavefront coherence time ($\tauO$) above the two ESO sites of Cerro Paranal 
(site of the Very Large Telescope - VLT) and Cerro Armazones (site selected for the European Extremely Large Telescope - E-ELT).
The final outcome of the study is to investigate the opportunity to implement an automatic system for the 
forecast of these parameters at the VLT Observatory at Cerro Paranal and at the E-ELT Observatory at Cerro Armazones.

The optical turbulence (OT) forecast is a crucial cornerstone for the feasibility of the ELTs: it is indeed fundamental to support all kind of AO facilities 
in an astronomical observatory and to perform the flexible-scheduling of scientific programs and instrumentation through the 
Service Mode. 
Most of the scientific programs, associated to the most challenging scientific goals, 
require, indeed, frequently excellent optical turbulence conditions to be carried out. 
The traditional queue system, that is based on the quality of the scientific program 
but that does not take into account the OT conditions, leads necessarily to a paradox: 
the higher is the scientific challenge of a scientific program, the lower is the probability 
to complete the program itself. From this we derive that the Service Mode, that takes into 
account the status of the OT beside to the quality of the scientific program, 
is mandatory to optimize the exploitation of the ELTs. 
However, to implement a Service Mode we need to forecast the OT to know 
its status at different delayed times $\Delta$T with respect to the time at which the prediction 
is performed. 
The optimization of the {\it use} of a ground-based facility has therefore serious 
implications on the final scientific impact of the facility itself. We note, moreover, that the cost 
of a night of observation is of the order of hundreds of KDollars. The implementation of an OT 
forecast system leads therefore to a not negligible rationalization of costs versus scientific feedbacks. 
We remind also that, for evident statistical reasons, the advantage of the Service Mode can be fully 
achieved only if most of the available observing time is scheduled in this mode. For all these reasons 
ELTs plan to have, in their baseline configuration, permanent instruments located at different focal 
stations. It has been estimated that, at the E-ELT, the typical time required to switch the beam from 
an instrument to another is of the order of 10-20 minutes \citep{ESO_RD6}. This is therefore the final minimal 
time-scale in terms of OT prediction that we can take as a reference. All these premises lead to the 
conclusion that our ability in forecasting the OT is fundamental for the success and the feasibility of the ELTs. 

%A few different methods/techniques have been tested for this purpose in the past: statistical methods, analytical methods, neural network methods, physical methods. 
%Most of these methods are based on strong approximations and/or empirical assumptions. 
%In some cases they can provide typical trend but not really a specific prediction. 
%A detailed description of the methods with their characteristics can be found in \citet{masciadri1999a} and in the review \citet{masciadri111}. 
The peculiar feature of the numerical approach, and more precisely of the mesoscale atmospheric models, is the possibility to describe a 3D map of 
the $\CN2$ (i.e. volumetric distribution of the optical turbulence in the troposphere) in a region 
around a telescope of the order of a few tens of kilometers at a time t$_{0}$ placed in the future. 
This is the unique method that aims at address directly the solution of the Navier-Stokes equation of the hydrodynamical flow above the Earth. 
It is therefore intrinsically the most appropriate to detect directly a rapid change of the atmospheric status.

The typologies of models that can be used to study thermodynamic evolution of the earth's atmosphere are: 
the General Circulation Models (GCM), the non-hydrostatic meso-scale models, the models for Large Eddy Simulations (LES) and the models for 
Direct Numerical Simulations (DNS). 
These models differ in their resolution and, as a consequence, also in the extent of the typical domain of the atmosphere that they can reconstruct. 
We deduce that each of these typologies of model is dedicated to resolve phenomena of different nature and that the numerical approach as well as the 
physics described inside each of these models is done following a completely different logic. 

The OT can not be resolved by the GCMs because the spatio-temporal fluctuation of the OT is much smaller (order of centimeter/meter) 
than the typical resolution of the GCM ($\sim$16~km). 
With the DNS the OT can be completely resolved but we can not forecast it because we definitely miss the link with initialization data i.e. with the 
evolution of the atmospheric flow at large spatial scales. 
The meso-scale models represent the right trade-off that permits to reconstruct the OT maintaining the link with the external spatio-temporal 
evolution of the atmospheric flow. 
The OT is completely parameterized in the meso-scale models. 
With the LES the OT is partially resolved and partially parameterized. In principle it is possible in perspective to use the LES to improve 
the resolution of the simulations if initialized with outputs coming from meso-scale models. 
It has been observed that, when the horizontal resolution increases and reaches values of $\Delta$X smaller than 10~km, 
the hydrostatic models show a tendency in distorting the spectrum of the gravity waves, particularly in proximity of mountain regions. 
For this reason, non-hydrostatic models are more suitable to reconstruct the atmospheric flow on such conditions.  

This is the first paper of a series aiming at collecting the main results we obtained in the context of MOSE.
In Section \ref{for_par} we present the parameters that we intend to reconstruct with such a model and the importance they cover in the context of the astronomical observations and in the observational scheduling. In Section \ref{model} we describe the novelties of MOSE with respect to the status of art and we briefly describe the model we use for this study: Meso-Nh plus the Astro-Meso-Nh package. In Section \ref{config} we describe the model configuration used for the MOSE project. In Section \ref{obs_data} we describe the whole set of measurements we used for the model validation. In Section \ref{vert_atm} we present the model validation in reconstructing the vertical stratification of the atmospherical parameters. In Section \ref{conc} we present the conclusions of the study.

\section{Forecasted parameters}
\label{for_par}
In the context of the MOSE project we focused our attention on the following  
parameters that are strictly related to the optimization 
of the AO systems and operational and observational programs scheduling activities: all the astroclimatic parameters ($\varepsilon$, $\thetaO$, $\tauO$) and the atmospheric parameters from which the OT mainly depends on. \\ \\
\noindent
{\bf - Surface temperature:} the knowledge of the temperature near the ground is extremely important to thermalize the dome and 
reduce/eliminate the dome seeing, the most important contribution of the seeing experienced on ground-based images. 
The seeing inside the dome can be very hardly modeled being that the turbulence develops inside a confined environment and the Kolmogorov theory 
is rarely respected under these conditions.
It has been proved \citep{racine91} that the dome-seeing is mainly determined by (T$_{m}$-T$_{i}$)$^{6/5}$ and 
(T$_{i}$-T$_{o}$)$^{6/5}$ where T$_{m}$ is the primary mirror temperature, T$_{i}$ is the temperature inside the dome and T$_{o}$ 
is the temperature outside the dome, near the ground. Once T$_{o}$ is forecasted, it is possible to tune T$_{i}$ and T$_{m}$ so as to minimize the 
temperature gradients and eventually eliminate the dome seeing. 
The temperature gradient near the ground is moreover extremely critical to well reconstruct the optical turbulence near the ground. It is fundamental for us to know how good or bad is the model performance in reconstructing this parameter to discriminate the potential discrepancies of the model. \\ \\
\noindent
{\bf - Surface wind speed:} the intensity of  the wind speed near the ground can strongly affect the AO performances because of vibrations 
induced by all critical structures such as primary and secondary mirrors (particularly critical are the adaptive secondaries). 
Vibrations produced by wind speed bursts are among the most annoying source of noise for these elements and the accuracy of the 
AO correction is directly proportional to the noise introduced by such a kind of vibrations. 
An efficient forecast of the wind speed intensity can usefully optimize the use of the most sophisticated AO techniques and even decide when it is better to avoid AO 
observations if conditions do not permit to achieve required conditions. 
Besides, the wind speed shear near the ground is one of the main cause for OT triggering together with the temperature gradient. The interest in studying the ability of the model in reconstructing this parameter is therefore similar to what already said for the surface temperature. \\ \\
\noindent
{\bf - Surface wind direction:} the wind direction is one of the atmospheric parameters that are more frequently and easily 
correlated to the local seeing characteristics. 
The prediction of the wind direction can therefore give a direct information on the probability to have bad or good seeing.  It becomes very important when the wind speed is strong because, under this condition, the effects of the wind speed on the adaptive secondary (vibrations) are strongly dependent on the wind direction.\\ \\
\noindent
{\bf - Vertical profiles on $\sim$ 20~km: potential temperature, wind speed intensity and direction and relative humidity:} 
such a stratification can be obtained, in principle, by GCMs. However, it has been proved 
\citep{masciadri01a,masciadri03,hagelin10}
that GCMs are not reliable in the low part of the atmosphere particularly in proximity of mountain regions because of their too low horizontal resolution 
and a too smooth orography. 
Moreover, it is also worth reminding that a meso-scale model can produce a continuous temporal evolution of the vertical stratification of these parameters 
with a temporal sampling that can be freely selected (in our study we use a temporal sampling of 2 min) while information provided by GCM is sampled only at synoptic hours (00:00, 06:00, 12:00 and 18:00 UT). The meso-scale model technique appears therefore as particularly interesting for systematic calculations of the temporal evolution of the  wavefront coherence time 
at astronomical observatories. 
Besides, the wind speed intensity shear (together with the potential temperature gradient) are the fundamental cause of the OT triggering. 
It is therefore important to assure us how the model reconstructs the vertical stratification of these parameters to be able to discriminate the potential causes of model discrepancies or failures. The potential temperature $\theta$ of an air parcel is defined as the temperature which the parcel of air would have if it were expanded or compressed adiabatically from its existing pressure and temperature to a standard pressure P$_{0}$ (generally taken as 1000 mb):

\begin{equation}
\theta =T(\frac{P_{0}}{P})^{R/c_{p}}
\end{equation}
where R is the gas constant R = 287 J$\cdot$K$^{-1}$$\cdot$kg$^{-1}$, c$_{p}$ is the specific heat capacity at constant pressure c$_{p}$ = 1004 J$\cdot$K$^{-1}$$\cdot$kg$^{-1}$, P is the pressure and T is the absolute temperature. It is preferable to study the stratification of $\theta$ instead of that of the absolute temperature T because $\theta$ is strictly related to the thermodynamic stability of the atmosphere that is one of the principal source triggering the optical turbulence. In the astronomical context we count already some studies aiming at characterizing the sites for their thermodynamic stability \citep{geissler2006,hagelin2008,hach2012}. \\ \\
\\ \\
{\bf - $\CN2$ profiles:}
The $\CN2$ profiles tell us how the turbulence is spatially distributed in the whole atmosphere, where are located the main turbulent layers during a night. This is a fundamental information for the AO tomography. The algorithm used for the $\CN2$ parameterization is the one described in \citet{masciadri1999a}. 
We refer the reader to the same paper for the detailed description of the theory connecting the dynamic turbulence with the optical 
turbulence and the algorithms used in the Meso-Nh model. 
In the context of the MOSE project we also tested a modified/new algorithm (new parameterization) in order to test which of the 
two ones provides better model performances. 
In a forthcoming paper the algorithm will be presented as well as the results obtained. \\ \\
\noindent
{\bf - Integrated astro-climatic parameters:} in the context of the MOSE project it has been established to investigate the three 
most important integrated astro-climatic parameters\footnote{We highlight that the Astro-Meso-Nh package includes also other parameters such as the scintillation rate and the spatial coherence outer scale.} :  the seeing ($\varepsilon$), the isoplanatic angle ($\theta_{0}$) and the wavefront coherence time ($\tau_{0}$):\

\begin{equation}
r_{0}=\left [ 0.423\left ( \frac{2\pi }{\lambda} \right )^{2} \cdot \int_{0}^{\infty }C_{N}^{2}(h)dh)\right ]^{-3/5}
\end{equation}
\begin{equation}
\varepsilon =0.98\frac{\lambda }{r_{0}}
\end{equation}
\begin{equation}
\theta _{0}=0.057\lambda ^{6/5}\left [ \int_{0}^{\infty } h^{5/3}C_{N}^{2}(h)dh)\right ]^{-3/5}
\end{equation}
\begin{equation}
\tau _{0}=0.057\lambda ^{6/5}\left [ \int_{0}^{\infty } \left | V(h) \right |^{5/3}C_{N}^{2}(h)dh)\right ]^{-3/5}
\end{equation}

The knowledge of the values of these parameters is critical to manage and optimize the AO systems. The seeing is the main discriminant permitting us to identify the temporal windows in which the AO systems can be mostly effective, $\tauO$ tell us how fast is the turbulence, $\thetaO$ tells us how much turbulence is present in the free atmosphere particularly in the high part of the atmosphere and it can be very useful to discriminate between AO observations to be done in narrow or wide field.

%%%%%%%%%%%%%%%%%%%%%%%%%%%%%%%%%%%
\section{Meso-Nh model and Astro-Meso-Nh package}
\label{model}

\subsection{Status of art}
Which are the MOSE scientific goals with respect to the status of art ? 
In 1999 \citep{masciadri1999a,masciadri1999b} it has been proved for the first time that a meso-scale model could reconstruct reliable $\CN2$ profiles.  
From a qualitative point of view, it has been proved that the shape of the profile could match the observed one, and from a quantitative point of view
 the observed and simulated integrated values of the turbulence present in the atmosphere were well correlated. 
In 2001 \citep{masciadri2001} a calibration method for the model has been proposed and the whole technique has been validated 
on a statistic of 10 nights comparing simulations with measurements taken from different instruments (balloons and Generalized SCIDAR) \citep{masciadri2004}. 
It has been proved that the dispersion of the seeing simulated and observed $\Delta$$\varepsilon_{model,GS}$ $\sim$ 25$\%$ 
was comparable to the dispersion obtained using different instruments $\Delta$$\varepsilon_{ballons,GS}$ $\sim$ 30$\%$. 
In a successive time the same technique has been applied in an autonomous way on a whole solar year at San Pedro M\'artir \citep{masciadri2006} carrying out a 
complete statistical analysis of all the astro-climatic parameters permitting us to achieve the first complete and homogeneous analysis of the 
seasonal variation investigation of the OT ever done so far. In 2011 \citep{hagelin2011} the model/technique has been validated using a richer 
statistical sample of GS measurements performed on 43 nights uniformly distributed along one year at Mt. Graham. At the same time it has been also proved that this technique can be efficient in discriminating sites characterized by substantial different turbulence features.
The Meso-Nh model has been applied to the internal Antarctic Plateau achieving two major conclusions: 
{\bf (1)} it has been proved that such a model is able to discriminate between the sites on the top of the summit of a plateau characterized 
by a very thin surface layer (Dome C, Dome A and South Pole) \citep{lascaux2009,lascaux2010,lascaux2011}; 
{\bf (2)} thanks to the numerical technique (Meso-NH model), the first estimates of the OT vertical stratification ever done in the free atmosphere at Dome A have been provided \citep{lascaux2011}.\\
\noindent
We finally remind that, the heart of the OT parameterization used in our study \citep{masciadri1999a} has been implemented by the colleagues of Mauna Kea \citep{cherubini2011} on the WRF mesoscale model for applications to the Mauna Kea Observatory in an operational configuration. An extended review on alternative methods for the OT forecasts can be found in \citet{masciadri2011}. We limit here to cases of an OT parameterized in a non-hydrostatical mesoscale model.

\subsection{MOSE scientific goals: novelties}
The MOSE project aims at overcoming two major limitations that are typically encountered in studies focused on the optical turbulence forecast with 
atmospheric models: {\bf (1)} the difficulty in having independent samples of measurements for the model calibration and model validation 
to estimate if and how the correlation between measurements and predictions on the validation sample changes with the increasing of the number of nights and to estimate if and how the statistical richness of the calibration sample affects the calibration itself; 
{\bf (2)} the difficulty in having a large number of simultaneous measurements done with different and independent 
instruments for the OT estimates (in particular vertical profilers). 
This project is performed with the non-hydrostatic mesoscale atmospherical model Meso-NH \citep{lafore98} joined with the Astro-Meso-NH package
for the calculation of the optical turbulence \citep{masciadri1999a,masciadri1999b} to perform the OT forecasts. 
An extended data-set of observations (meteorological parameters and optical turbulence) have been 
considered in the project. 
We took advantage of measurements obtained in the context of the site selection 
for the TMT (American study) at Cerro Armazones and on measurements taken routinely 
during the last decade and/or in a dedicated site testing campaign (PAR2007) at Cerro Paranal (details in Section \ref{obs_data}). 

\subsection{Tool of investigation}

All the numerical simulations of the nights presented in this study have been performed with the non-hydrostatical mesoscale numerical weather model
Meso-NH\footnote{$http://mesonh.aero.obs-mip.fr/mesonh/$} \citep{lafore98}. 
The model has been developed by the Centre National des Recherches M\'et\'eorologiques (CNRM) and Laboratoire d'A\'ereologie (LA) 
de l'Universit\'e Paul Sabatier (Toulouse). 
The Meso-NH model can simulate the temporal evolution of three-dimensional meteorological
parameters over a selected finite area of the globe.
The system of hydrodynamic equations is based upon an anelastic formulation allowing for an effective filtering of acoustic waves.
It uses the Gal-Chen and Sommerville \citep{gal75} coordinates system on the vertical and the C-grid in the formulation of
Arakawa and Messinger \citep{arakawa76} for the spatial digitalization.
It employs an explicit three-time-level leap-frog temporal scheme with a time filter \citep{asselin72}.
It employs a one-dimensional 1.5 turbulence closure scheme \citep{cuxart00}.
For this study we use a 1D mixing length proposed by Bougeault and Lacarr\`ere \citep{bougeault89}.
The surface exchanges are computed using ISBA (Interaction Soil Biosphere Atmosphere) module \citep{noilhan89}.

The optical turbulence and derived parameters are not an intrinsic part of the Meso-NH model 
but it has been developed by our team in an independent package. 
The package (called Astro-Meso-NH), includes the algorithms 
for the \CN2 parameterization and all the other integrated astro-climatic parameters 
(seeing, isoplanatic angle, wavefront coherence time, scintillation rate, spatial coherence outer scale,..). 
The first version of the code has been presented in \citep{masciadri1999a}. 
Since then the code have been improved along the years in terms of flexibility providing a set of further outputs and 
permitting us to carry out different scientific studies mainly addressing the reliability of this technique.

Both the Meso-Nh code as well as the Astro-Meso-NH code for the optical turbulence are parallelized with OPEN-MPI-1.4.3.
The model can therefore be run on local workstations as well as on the High Performance Computing Facilities (HPCF) cluster of the 
ECMWF, in parallel mode so as to gain in computing time.

\begin{figure*}
\begin{center}
\begin{tabular}{c}
\includegraphics[width=\textwidth]{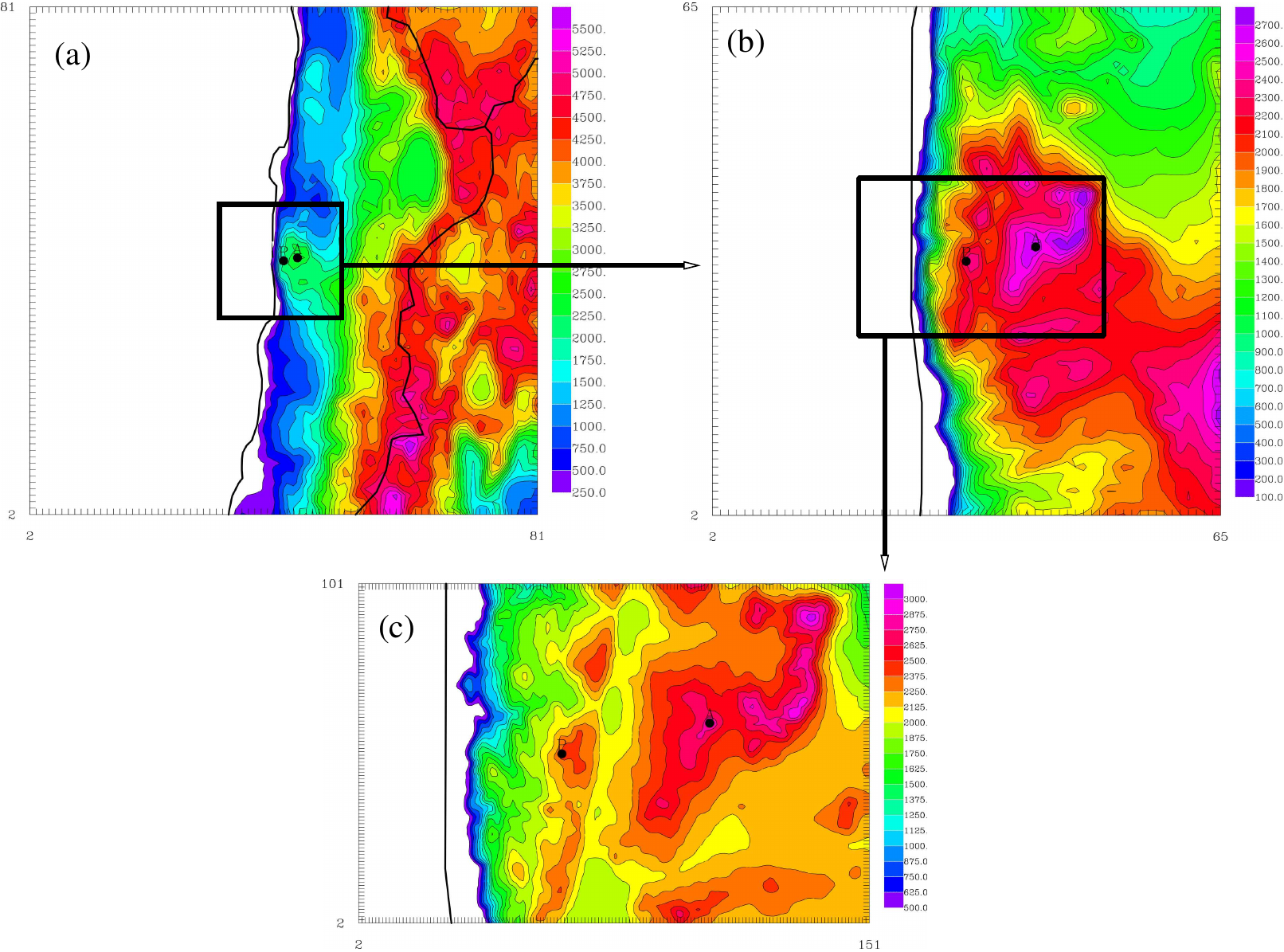}
\end{tabular}
\end{center}
\caption{\label{fig:grid-nesting_3mod} Orography (altitude in m) of the region of interest
as seen by the Meso-NH model (polar stereographic projection)
for all the imbricated domains of the grid-nested standard configuration.
{\bf (a)} Domain 1 (orographic data from GTOPO),
{\bf (b)} Domain 2 (orographic data from GTOPO),
{\bf (c)} Domain 3 (orographic data from ISTAR),
A dot stands for Cerro Armazones and P dot stands for Cerro Paranal.
See Table~\ref{tab:gn_config1} for the specifications of the domains (number of grid-points, domain extension, horizontal resolution).}
%Standard grid-nesting configuration (3 imbricated domains)
%\caption[grid-nesting_3mod]{\label{fig:grid-nesting_3mod} Standard grid-nesting configuration (3 imbricated domains)
% with (a) the outermost domain with a horizontal resolution $\Delta$X = 10 km; (b) the second domain with a
%horizontal resolution $\Delta$X = 2.5 km; (c) the innermost domain with a horizontal resolution $\Delta$X = 500 m. Units in m.
%More details are reported in table \ref{tab:gn_config1}.}
\end{figure*}

\begin{table}
\caption{Meso-NH model standard grid-nesting configuration.
In the second column the number of horizontal grid-points, in the third column the domain extension and
in the fourth column the horizontal resolution $\Delta$X.}
\begin{center}
\begin{tabular}{|c|c|c|c|}
\hline
Domain & Grid   & Domain size & $\Delta$X  \\
       & Points & (km)        & (km)  \\
\hline
Domain 1 &  80$\times$80  & 800$\times$800 & $\Delta$X = 10\\
Domain 2 &  64$\times$64  & 160$\times$160 & $\Delta$X = 2.5\\
Domain 3 & 150$\times$100 &  75$\times$50  & $\Delta$X = 0.5\\
\hline
\end{tabular}
\label{tab:gn_config1}
\end{center}
\end{table}

\begin{table}
\caption{Meso-NH model high resolution grid-nesting configuration.
In the second column the number of horizontal grid-points, in the third column the domain extension and
in the fourth column the horizontal resolution $\Delta$X.}
\label{tab:gn_config2}
\begin{center}
\begin{tabular}{|c|c|c|c|}
\hline
Domain & Grid   & Domain size & $\Delta$X \\
       & Points & (km)        &  (km)  \\
\hline
Domain 1 &  80$\times$80  & 800$\times$800 & $\Delta$X = 10   \\
Domain 2 &  64$\times$64  & 160$\times$160 & $\Delta$X = 2.5 \\
Domain 3 & 150$\times$100 &  75$\times$50  & $\Delta$X = 0.5   \\
Domain 4 & 100$\times$100  & 10$\times$10  & $\Delta$X = 0.1   \\
Domain 5 & 100$\times$100 &  10$\times$10  & $\Delta$X = 0.1   \\
\hline
\end{tabular}
\end{center}
\end{table}
%%%%%%%%%%%%%%%%%%%%%%%%%%%%%%%%%%
\section{Model configuration}
\label{config}

The geographic coordinates of Cerro Paranal are (24$^{\circ}$37'33.117" S, 70$^{\circ}$24'11.642 W) and those of Cerro Armazones are (24$^{\circ}$35'21" S, 70$^{\circ}$11'30" W).
The grid-nesting technique \citep{stein00}, employed in our study, consists of using different imbricated domains
of the Digital Elevation Models (DEM i.e orography) extended on smaller and smaller surfaces, with increasing horizontal
resolution but with the same vertical grid.
Two different grid-nesting configurations have been employed.
The standard configuration includes three domains (Fig.~\ref{fig:grid-nesting_3mod} and 
Table~\ref{tab:gn_config1}) and the innermost resolution is $\Delta$X~=~500~m.
The second configuration is made of five imbricated domains, the first same three as the previous configuration, and other two
centered at both Paranal and Armazones sites, with a horizontal resolution of $\Delta$X~=~100~m (Table~\ref{tab:gn_config2}). 
A forthcoming paper is dedicated
to the analysis of simulations obtained with such a high horizontal resolution DEM (Lascaux et al. 2013, submitted to MNRAS).
One can notice that using these configurations, we are able to do the forecast at both sites simultaneously.
The orographic DEMs we used for this project are the GTOPO\footnote{$http$:$//www1.gsi.go.jp/geowww/globalmap$-$gsi/gtopo30/gtopo30.html$}
with an intrinsic horizontal resolution of 1~km (used for the domains 1 and 2) and the ISTAR\footnote{Bought by ESO at the ISTAR Company
- Nice-Sophia Antipolis, France. The method is based on couple of stereoscopic images of the same location taken with different angles, obtained by SPOT satellites.} with an intrinsic horizontal resolution of 0.5~km (used for the domain 3, 4 and 5).
It is important to note that we used a "two way" grid-nesting that means that 
the atmospheric flow has a mutual interaction at the interface of each couple of domains: the larger domain (called 'father') and the innermost smaller domain (called 'son'). This allows that the atmospheric flow in the smaller domain is constantly in a thermo-dynamic equilibrium with the flow at larger scales. This guarantees the propagation of the gravity waves that can be triggered by the interaction of the atmospheric flow on the Chilean coast and reach the atmosphere above Cerro Paranal or Cerro Armazones in the free atmosphere.  
Along the z-axis we have 62 levels distributed as follows: a first vertical grid point equal to 5~m,
a logarithmic stretching of 20~$\%$ up to 3.5~km above the ground, and an almost constant vertical grid size of $\sim$600~m up to 23.8~km.\\
All the vertical profiles extracted from the model computations are available every 2~min.
All the computed meteorological parameters near the surface [0-30]~m a.g.l. are available at every time step of the innermost domain, which is 
$\Delta$t~=~3~s for the $\Delta$X~=~500~m configuration and $\Delta$t~=~1~s for the $\Delta$X~=~100~m configuration.

%%%%%%%%%%%%%%%%%%%%%%%%%%%%%%%%%%
\section{Whole observations data-set}
\label{obs_data}
\noindent
{\bf Atmospherical parameters near the surface [0-30]~m.}\\
At Paranal, observations of meteorological parameters near the surface come from an automated weather station (AWS) and a 30~m high mast including a number of sensors placed at different heights. Both instruments are part of the VLT Astronomical Site Monitor \citep{sandrock99}. Absolute temperature data are available at 2~m and 30~m above the ground. Wind speed data are available at 10~m and 30~m above the ground. At Armazones, observations of the meteorological parameters near the ground surface come from the Site Testing Database \citep{schoeck09},  more precisely from an AWS and a 30~m tower (with temperature sensors and sonic anemometers). Data on temperature and wind speed are available at 2~m, 11~m, 20~m and 28~m above the ground. At 2~m (Armazones) temperature measurements from the AWS and the sonic anemometers are both available but we considered only those from the tower (accuracy of 0.1$^{\circ}$C)\citep{skidmore07}. Those from the AWS are not reliable because of some drift effects (T. Travouillon, private communication). Wind speed observations are taken from the AWS (at 2~m) and from the sonic anemometers of the tower (at 11~m, 20~m and 28~m). The outputs are sampled with a temporal frequency of 1 minute. \\ \\
\noindent
{\bf Atmospherical parameters: vertical stratification}\\ 
At Paranal we had access to 50 radio-soundings (vertical distribution of the meteorological parameters in the $\sim$20~km above the ground) launched above this site in the context of an intense site testing campaign for the water vapor characterization \citep{cha11} and covering 23 nights in 2009, 11 nights in summer and 12 in winter time. In a subsample of these nights (16), a few radio-soundings (two or three) have been launched at different times during the same night.  The radiosoundings used Vaisala radiosondes (RS92 - SGP) with an accuracy of 0.5 $^{\circ}$C for the temperature, an accuracy of 5\% for the
relative humidity, an accuracy of 0.15 m$\cdot$s$^{-1}$ for the wind speed and an accuracy of 2$^{\circ}$ for the wind direction.
They also have a positioning uncertainty on the vertical direction of 20~m. A few more details will be provided in Section \ref{gen}).
\\ \\
\noindent
{\bf Optical turbulence}\\ 
Observations of the optical turbulence at Paranal, related to the Site Testing Campaign of November-December 2007 \citep{dali10}, come from a Generalized SCIDAR (GS called CUTE-Scidar III developed by the Istituto de Astrofisica de Canarias (IAC) team) \citep{vazquez2008}, a Differential Image Motion Monitor (DIMM) \citep{sarazin1990} and a Multi Aperture Scintillation Sensor (MASS) \citep{tokovinin2003}. The GS measurements have been recently re-calibrated \citep{masciadri2012} to correct intrinsic errors of the GS due to the normalization of the auto-covariance of the scintillation maps by the auto-correlation of the pupil image. Optical turbulence measurements at Cerro Armazones come from a DIMM and a MASS that have been used for the TMT site selection campaign \citep{schoeck09}.

%%%%%%%%%%%%%%%%%%%%%%%%%%%%%%%%%%
\section{Atmospheric parameters vertical stratification: model vs. observations}
\label{vert_atm}

%%%%%%%%%%%%%%%%%%%%%%%%%%%%%%%%%%
\subsection{General Information}
\label{gen}
{\bf Site}: Cerro Paranal;\newline
{\bf Instruments}: ballon radiosoundings;\newline
{\bf Parameters investigated}: wind speed, wind direction, potential temperature, relative humidity;\newline
{\bf Number of nights}: 23 nights in 2009 (12 during winter; 11 during summer);\newline
{\bf Atmospheric region investigated}: [0-20]~km;\newline

The balloons have been launched close to the synoptic hours (00:00 UT, 06:00 UT, and/or 12:00 UT) in 23 different nights (see list of nights and hour of launch in Annex \ref{annex_a}). These nights 
were used to investigate the performances of the model 
in reconstructing the vertical distribution of the meteorological parameters. To do this,  we performed 23 simulations (1 for every night), each starting at 
18:00~UT of the day before and ending at 14:00~UT of the simulated day (for a total duration of 20 hours). 
We then compared observations from the radio-soundings with the simulations performed by the
model in the range [3 - 21] km a.s.l., because the first 400-500 meters above the ground (h~=~2634 m - Paranal summit)
constitute, a sort of 'grey zone' in which it is meaningless to compare any quantitative estimates 
because of many different reasons. 
Among others: {\bf (a)} the orographic map of
the innermost domain has an intrinsic $\Delta$h $\simeq$ 156~m with respect to the real summit due to the natural smoothing
effect of the model horizontal interpolation of the DEM;
{\bf (b)} the radio-soundings have been launched at around
50 m below the summit. It should be meaningless to compare the observed and simulated values at the summit
ground height because in one case (model) we resolve friction of the atmospheric flow near the ground, in the other (measurements) no;
{\bf (c)} we have to take care about an uncertainty $\Delta$h of around 50 m in the identification of the zero point (h$_0$)
probably due to an unlucky procedure performed during the radio-soundings launches on the zero point
setting. This uncertainty has basically no more effects above a few hundreds of meters above the ground because the
atmospherical parameters we are studying are affected by phenomena evolving at larger spatial scales at these heights. 
We decided therefore to treat data only above roughly 500 m from the summit to overcome these uncertainties.
For the model,  we tok the averaged vertical profiles simulated by the model in around one hour from the time in which the radio-sounding has been launched.
We considered, indeed, that the balloon is an in-situ measurement and a balloon needs around 1 hour to cover 20~km from the ground moving up in
the atmosphere with a typical vertical velocity of $\sim$6~m$\cdot$s$^{-1}$.

To estimate the statistical model reliability in reconstructing the main meteorological parameters we used the averaged values plus
two statistical operators: the bias and the root mean square error (RMSE) defined as:
\begin{equation}
BIAS = \sum_{i=1}^{N}\frac{Y_i-X_i}{N}
\label{eq:bias}
\end{equation}
\begin{equation}
RMSE = \sqrt{\sum_{i=1}^{N}\frac{(Y_i-X_i)^2}{N}}
\label{eq:rmse}
\end{equation}
where $X_i$ are the individual observations, $Y_i$ the individual simulations parameters calculated at the same time and $N$ is
the number of times for which a couple ($X_i$,$Y_i$) is available with both $X_i$ and $Y_i$ different from zero. The bias and the RMSE take into account all the systematic and statistical errors. Starting from the bias and the RMSE it is possible to retrieve the bias-corrected RMSE:
\begin{equation}\begin{split}
\sigma=\sqrt{\sum_{i=1}^{N}\frac{[(X_i-Y_i)-(\overline{X_i-Y_i})]^2}{N}}\\ =\sqrt{RMSE^2-BIAS^2}
\label{eq:sigma}
\end{split}\end{equation}

At the same time, due to the fact that we are interested in investigating the model ability in forecasting a parameter 
night by night and not only in characterizing it, for a few parameters we also calculated the correlation observations/simulations 
night by night and not only in statistical terms. This second goal is much more challenging in terms of score of success for evident reasons.

%%%%%%%%%%%%%%%%%%%%%%%%%%%%%%%%
\begin{figure*}
\begin{center}
\begin{tabular}{c}
\includegraphics[width=0.95\textwidth]{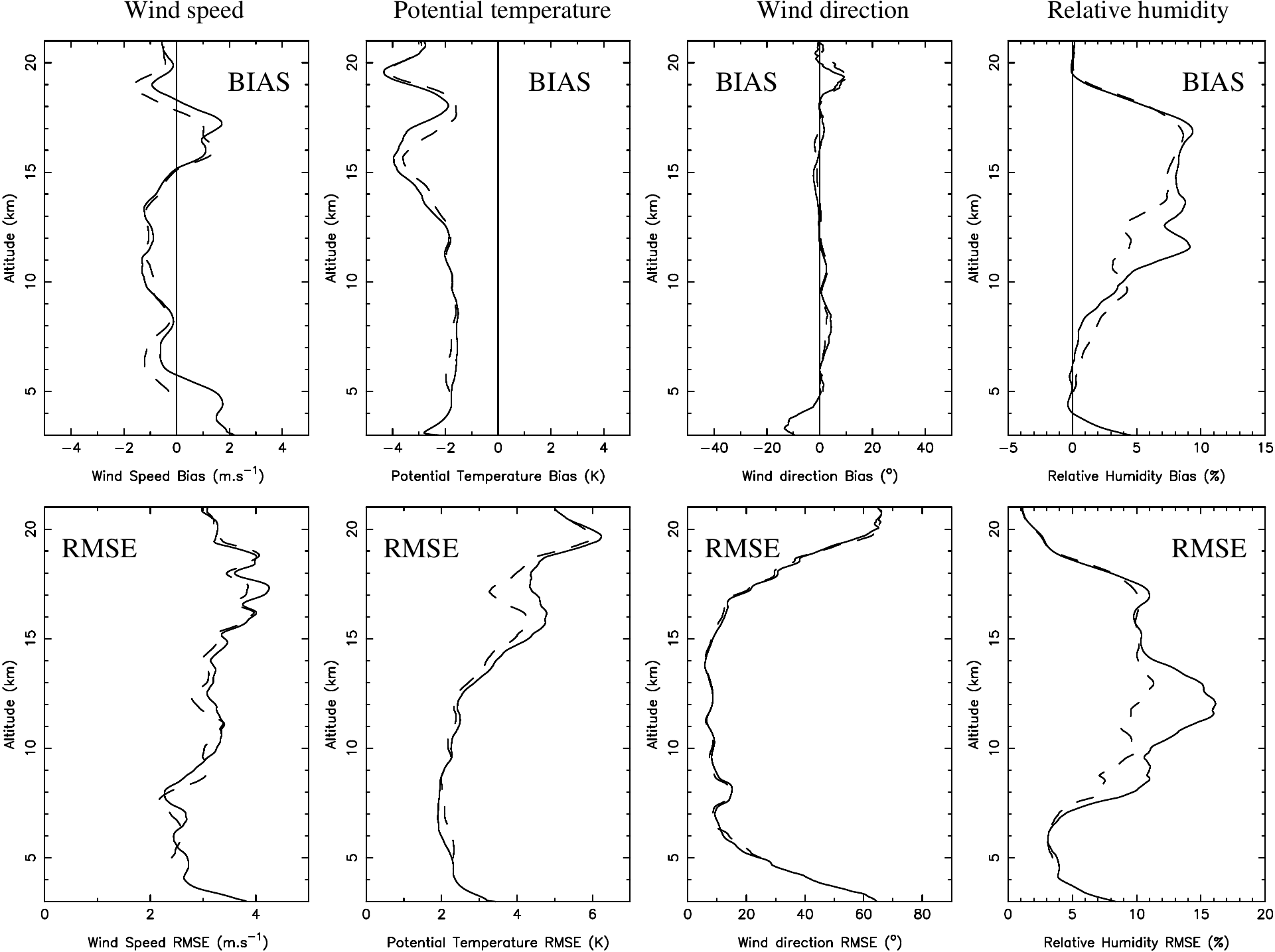}\\
\includegraphics[width=0.95\textwidth]{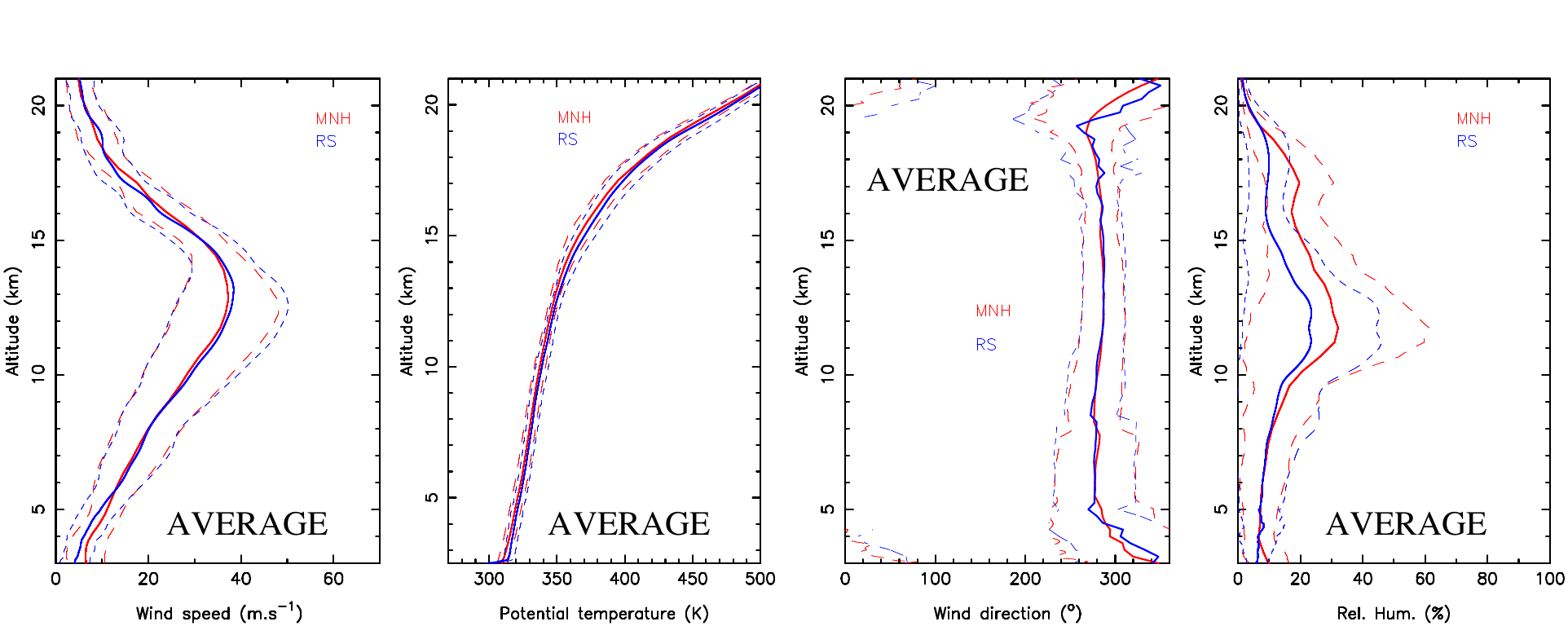}
\end{tabular}
\end{center}
\caption{First and second raws: Bias and RMSE (Meso-NH - Observations) and average of wind speed, potential temperature, wind direction and relative humidity (thin line). Bias and RMSE (ECMWF analyses - Observations) reported for h $>$ 5~km (see text) (dashed line). 
Data and model profiles interpolated on a 5 m-vertical grid; then bias and RMSE are computed for each interpolated levels; finally a moving average over 1 km 
is applied on the resulting profiles. In the bottom raw, dashed lines represent the standard deviation, full thin lines are the average. NB: for the wind direction average profile we do calculate the moving average. On the y-axis is reported the height above sea level. }
\label{fig:Average_Bias_and_RMSE_rs}
\end{figure*}

\subsection{Overall statistical model performances}

In this section we consider the overall statistics involving the full sample of 50 radiosoundings.
Fig.~\ref{fig:Average_Bias_and_RMSE_rs} shows the average vertical profiles, the bias and the RMSE calculated on this sample of 50 radiosoundings, 
of the wind speed, the potential temperature, the wind direction, and the relative humidity.
The bias contains information on systematic model errors. The RMSE contains information on the statistical errors plus the systematic errors. 
To elaborate theses figures, first observations and model profiles have been interpolated on a 5~m vertical grid, then average, bias and RMSE have been 
computed for each interpolated levels, and finally a convolution with a 1~km width has been applied on the resulting profiles.\\
Looking at Fig.~\ref{fig:Average_Bias_and_RMSE_rs} between 5~km a.s.l. and 18~km a.s.l. the model reconstruction of the wind direction shows a bias within just a few degrees and a RMSE within 10$^{\circ}$. 
Below 5~km a.s.l. (where the orographic effects are more evident) the bias is of the order of $\sim$20$^{\circ}$ with a RMSE that can reach 60$^{\circ}$. 
This means that, night by night, in the vertical slab [3 - 5] km a.s.l.~=~[0.5 - 2.5] km a.g.l., we 
can have a discrepancy of the order of a few tens of degrees. It is worth highlighting that at these heights the orographic effects are much more evident than at higher heights and they introduced disturbing effects particularly on wind direction and wind speed. Also it is true that measurements are done by balloons that are drifted horizontally during their flight that can introduce further sources of uncertainties for the wind speed in this vertical slab. In Section \ref{indiv} a more detailed discussion on this last issue will be presented. This is to say that such larger discrepancies at these heights are not necessarily due to a wrong estimate by the model. The larger values of the RMSE above $\sim$18~km is due, highly probably, to the fact that the wind speed sharply decreases at those heights. As a general rule, the wind direction is much more easily reconstructed when the wind speed is large.\\
For the relative humidity the bias is basically never larger than 10$\%$ all along the 20~km. 
The largest discrepancy (bias of 10$\%$) of simulations with respect to measurements is observed at the jet stream level where the RMSE can reach 15$\%$.
Such a satisfactory result has been obtained in spite of the fact we used a cheap scheme (in terms of CPU cost) for the relative humidity. 
That was possible because of the dryness of the region and because the relative humidity is not a parameter from which the optical turbulence is mainly driven. 
Such a solution permits faster simulations. 
The small bump at the jet-stream level (Fig.~\ref{fig:Average_Bias_and_RMSE_rs}, bottom right), highly probably due to the humidity coming from the close ocean,  tells us that the model reconstructs a little more humidity than what observed. Considering that General Circulation Models provide (as we will see later) similar results at these heights, we think that this excess of relative humidity is due to the initialization data that slightly overestimate the relative humidity at these heights.\\
%The model shows also a very good performances in reconstructing the potential temperature. 
For the potential temperature we observe a very small bias of $\sim$ 2$^{\circ}$C from the ground up to around 13~km a. s. l. 
Above 13~km a.s.l., where the potential temperature slope is steeper and steeper, the bias can reach up to 4$^{\circ}$C. \\
For the wind speed we have a bias of around 1~m$\cdot$s$^{-1}$ and a RMSE within 3 m$\cdot$s$^{-1}$ in the [5 - 15] km a.s.l. range. 
Above 15~km a.s.l. and in the [3 - 5] km a.s.l. range the bias reaches a value of 2~m$\cdot$s$^{-1}$. 
We conclude therefore that for the four atmospherical parameters the Meso-Nh model performances are very satisfactory.\\

By comparing results obtained with the Meso-NH model and the ECMWF\footnote{European Centre for Medium Range Weather Forecasts} analyses coming from the General Circulation Models (GCM) (see dashed lines in Fig.\ref{fig:Average_Bias_and_RMSE_rs}) we can see that the bias and the RMSE between Meso-Nh and observation and GCM and observations are basically the same in the region where the GCMs are reliable (i.e. above the first kilometers above the ground\footnote{There is no a precise rule to define a threshold. We considered 5~km a.s.l. just because it seems to be the height at which the orographic effects start to be less important.}). Only for the relative humidity, at the jet-stream level, the ECMWF shows a slightly weaker RMSE (a maximum of 12$\%$ instead of 17$\%$). However this is not necessarily due to an intrinsic problem of the Meso-Nh model but it can be due to the fact that, as we have previously said, a cheap scheme for the water has been used. The GCMs are applied to the whole Earth and they provide a complete 3D description of all the classical atmospherical parameters at synoptic hours (00:00, 06:00, 12:00 and 18:00) UT with a horizontal resolution of  25~km\footnote{This is the horizontal resolution for data of the 2009. ECMWF analyses have at present a horizontal resolution of 16~km}. A comparison of Meso-Nh versus GCMs performances permits us to retrieve insight on the origin of the residual discrepancies, at least in the regions of the atmosphere where this comparison is performed and it is meaningful. Being that the behavior of the two models  is very similar, this tells us that the residual biases and RMSEs we described so far are generated, highly probably, by initial conditions and not by the mesoscale model itself (numerical schemes, physical packages, etc.). A dedicated discussion on what happens in proximity of the vertical slab [3 - 5] km a.s.l. for the wind speed will be done in Section \ref{indiv}.\\ 
We note, finally, that the same calculation shown in Fig.\ref{fig:Average_Bias_and_RMSE_rs} has been done following two further different strategies. We considered:{\bf (a)} instantaneous model values at the round hour close to the balloon launch; {\bf (b)} instantaneous model values at the exact hour of the balloon launch. Results obtained are totally equivalent to those shown in Fig.\ref{fig:Average_Bias_and_RMSE_rs}. We think therefore the statistical results are quite solid in this respect.

%%%%%%%%%%%%%%%%%%%%%%%%%%%%%%%%%
\subsection{Individual nights model performances}
\label{indiv}

\begin{figure*}
\begin{center}
\begin{tabular}{c}
\includegraphics[width=14cm]{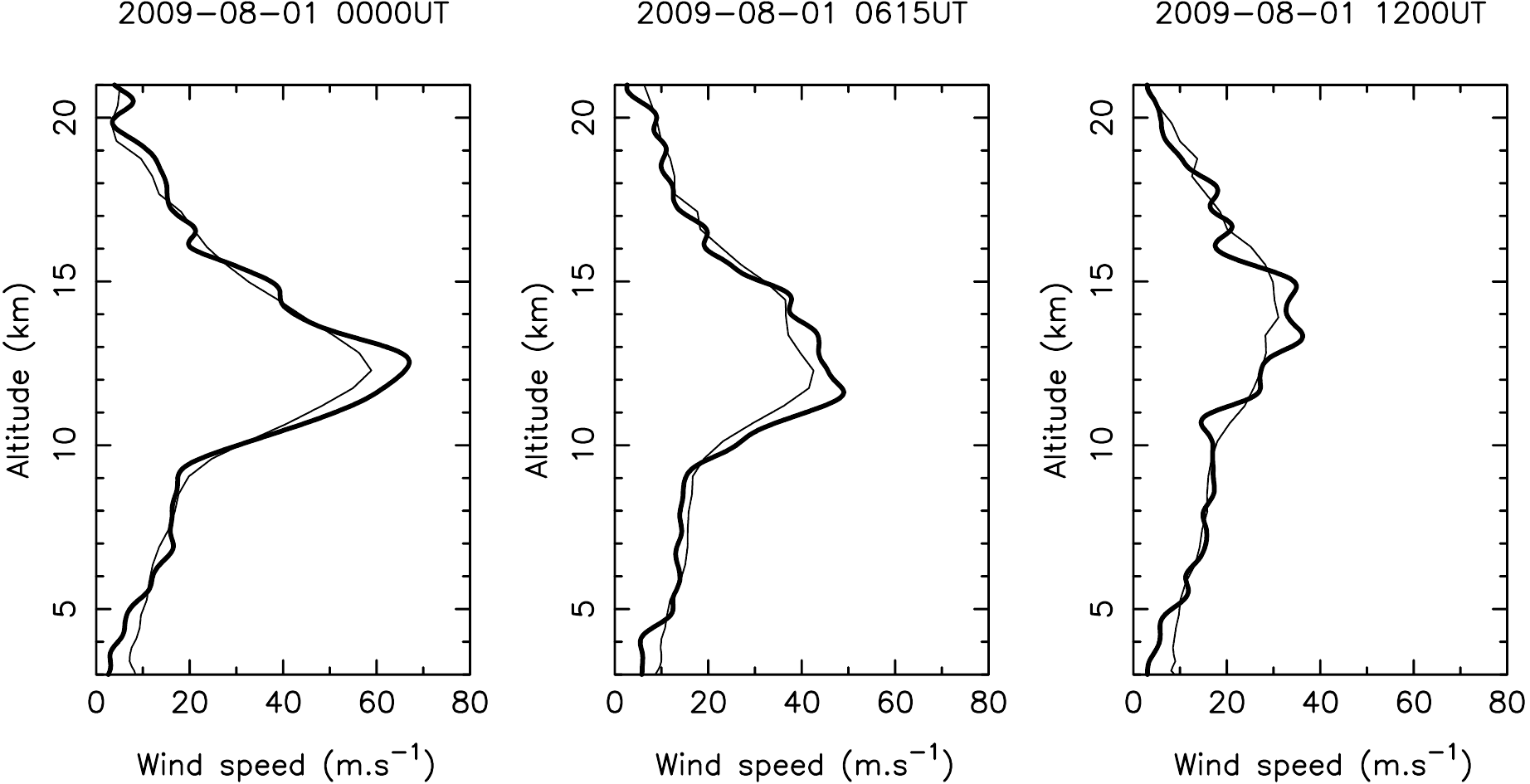}
\end{tabular}
\end{center}
\caption {Comparison of the wind speed intensity observed (radio-soundings: thick line) and simulated (Meso-NH model: thin line) at three different instants (00:00, 06:15, 12:00 UT) during the same night: 1/8/2009 above Cerro Paranal.}
\label{fig:evolt_rs} 
\end{figure*}

A comparison (observations/simulations) was performed night by night and at each time for which a radio-sounding was available for the wind speed. 
This parameter is particularly interesting in the context of the astronomical application to support AO facilities and manage observational scheduling of scientific programs. The wind speed retrieved by Meso-Nh might have multiple uses in this context. One could be interested, for example, {\bf (1)} in simply forecasting the wind speed to know in advance the temporal evolution of the wind speed all along the night for whatever general application; {\bf (2)} in using the forecasted wind speed joint with the forecasted $\CN2$ profiles (by the model itself) to forecast the wavefront coherence time $\tau_{0}$ in the next night; {\bf (3)} in using the forecasted wind speed joint with observed $\CN2$ in real time to calculate real time $\tau_{0}$. Indeed Meso-Nh can provide the wind speed vertical stratification with basically whatever temporal sampling. This is just a not necessarily exhaustive list of applications that might have some relevance in application to the ground-based astronomy supported by AO. \\
%The comparison revealed an excellent agreement model/observations in basically all the 50 cases studied. 
Fig.~\ref{fig:evolt_rs} shows the comparison of the wind speed observed and simulated in three different instants (00:00, 06:15 and 12:00 UT) of the same night (1/8/2009). We selected this night because it permits us to put in evidence in a clear way a specific model ability. We remind that the model has 62 vertical levels with a typical vertical resolution of $\sim$ 600~m above 3.5~km a.g.l.; the balloons have a typical vertical resolution of $\sim$ 6~m and they represent, moreover, instantaneous values. For the applications we have just cited we are obviously not interested in reconstructing the instantaneous high-spatial frequency structures of the wind speed (put in evidence by the balloons) but the averaged structures at large spatial scale. We remind, by the way, that the observed $\CN2$ profiles provided by instruments based on remote sensing principles have a vertical resolution of the order of 1~km (if we use the Generalized SCIDAR) or larger scales ($\Delta$h$\sim$0.5$\cdot$h) if we use the MASS. To compare the observed and simulated wind speed we perform therefore a convolution with a width of 1~km to filter out the fluctuations at high spatial frequency.  

Looking at Fig.\ref{fig:evolt_rs} we can note that, in spite of the fact that the observed wind speed strongly modifies its values all along 
the night at different heights, the model is able to reconstruct these changes and the observed wind speed values in the three different times in an impressive way. We are dealing of course about values at large spatial scale. 
In particular, the lowering of the intensity of the wind speed at the jet-stream level (it passes from 60-70~ms$^{-1}$ to 30-40~ms$^{-1}$) is well reconstructed by the model during the night. Being we are treating individual balloons flights we can not calculate a real statistical estimation on the individual nights. If we compare the simulated wind speed with an instantaneous measurements as in Fig.\ref{fig:evolt_rs} (thing that is not formally completely correct because we should considered an average of instantaneous measurements) we retrieve  that the largest relative error is 14 $\%$at the jet-stream level peak. A similar satisfactory behavior of the model is observed in basically all the 50 cases studied (see Annex \ref{annex_b}). Figures in Annex show just in a few cases and in isolated thin vertical slabs some variations reaching a relative error up to 20-25$\%$. We have no cases in which the model provides a total unreliable wind speed profile.\\

This definitely guarantees us the reliability of a tool (the Meso-NH mesoscale model) in reconstructing the temporal evolution of the vertical distribution of the wind 
speed V(h,t) during a whole night. This is a fundamental ingredient (beside to the vertical profiles of the optical turbulence $\CN2$(h,t)) 
to be used for the calculation of the temporal evolution of the wavefront coherence time $\tauO$(t):\\

\begin{equation}
\tau _0 (t) = 0.057 \cdot \lambda ^{6/5} (\int\limits_0^\infty  {V(h,t)^{5/3} }  \cdot C_N^2 (h,t)dh)^{ - 3/5} 
\end{equation}
\begin{figure*}

\begin{center}
\begin{tabular}{c}
\includegraphics[width=0.80\textwidth]{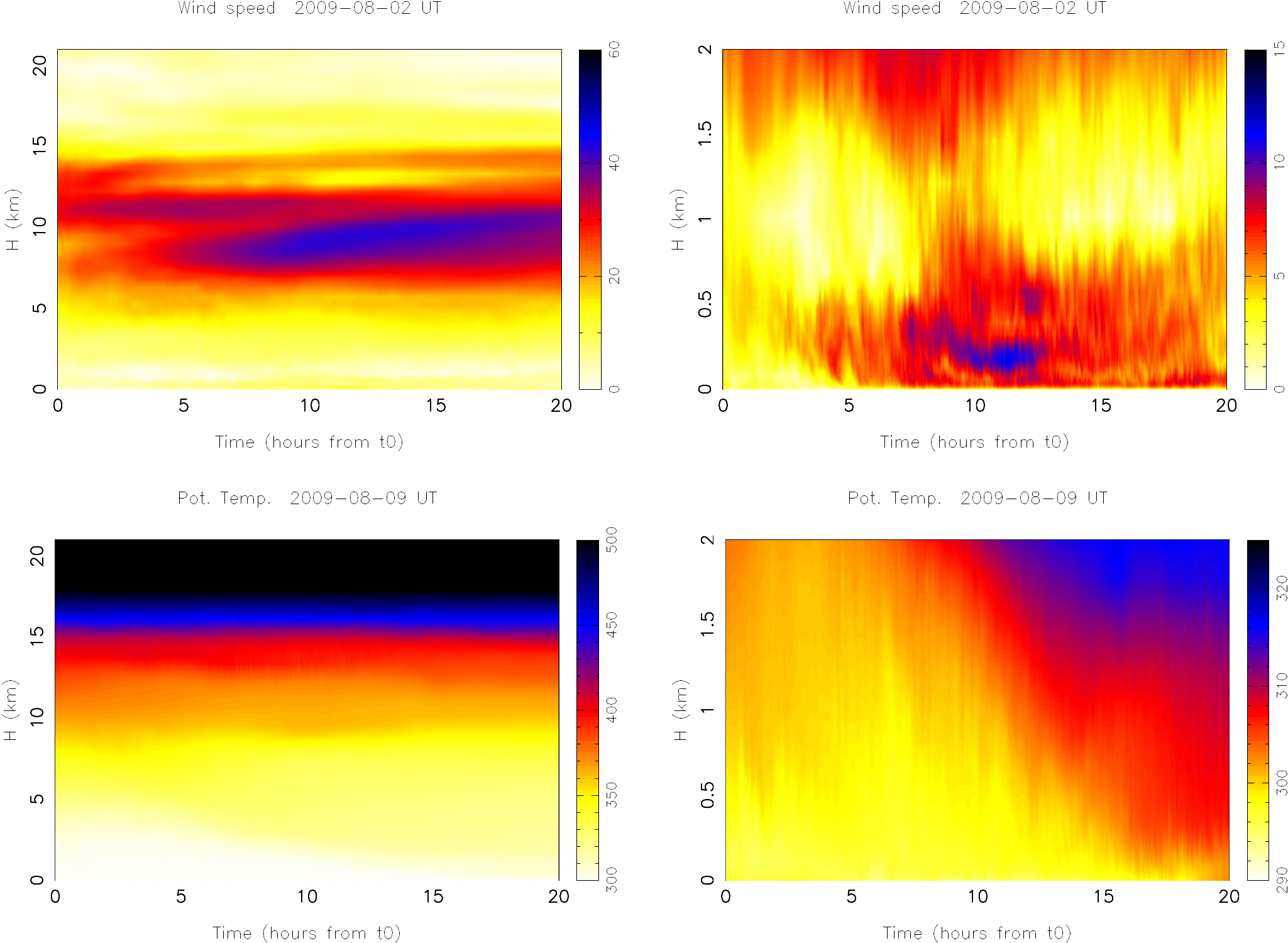}
\end{tabular}
\end{center}
\caption{Temporal evolution of the wind speed (top) and potential temperature (bottom) vertical distribution, 
calculated at the grid point of Paranal and extended along 21~km (left) and 2~km (right) from the ground. 
The simulation starts at t$_{0}$~=~18 UT and lasts 20 hours. The local night (20:00 - 05:00 LT) corresponds to the interval (6 - 15) on the x-axis.}
\label{fig:evolt_mnh} 
\end{figure*}
 
Fig.~\ref{fig:evolt_mnh} shows the temporal evolution of the wind speed and the potential temperature 
provided by the Meso-NH model during two different nights. 
In this example we can appreciate the intrinsic level of the temporal variability of both parameters at different heights above the ground during the night. 
This is far from being negligible and it tells us how much is important to be able to provide a continuum prediction over the time of these parameters.
In other words, the mesoscale predictions provide us a complete information (temporal evolution of the meteorological parameter) 
with respect to the estimations coming from the GCMs that provide outputs only at synoptic hours 
and that miss the information in between the synoptic hours. The GCMs, moreover, are not reliable in the low part of the atmosphere mainly for problems related to the low model horizontal resolution that affects the friction phenomena between the atmosphere and the ground.
All this tells us which is the invaluable utility of a mesoscale model for the prediction of the $\tauO$ and also for the calculation of the $\tauO$ in real time using the $\CN2$ from a vertical profilers and the wind speed from the Meso-Nh model. 
In particular, our results indicate that the wind speed retrieved from the mesoscale model is, at present, 
the cheapest and the most practical solution to calculate the temporal evolution of the wavefront coherence time $\tauO$(t). We remind that, at present, there are no practical automatic 'monitors' that are able to provide autonomous and systematical $\tauO$(t) measurements extended on the whole 20 km above the ground. Considering the reliability of the mesoscale models in reconstructing the wind speed profiles, at our opinion, the development of such monitors is not a priority, at least for the applications we have described. It is in principle possible to retrieve the wind speed profiles by a Generalized SCIDAR \citep{avila2006} (or even the derived technique of the Stereo-SCIDAR \citep{osborn2013}) but this requires a telescope of size larger than 1 m and it is not therefore very useful for systematic and autonomous monitoring of the wind speed. For this reason these instruments are more useful, at our opinion, for dedicated experiments.\\ \\
A few more words are suitable to comment the wind speed in the vertical slab [3 - 5] km a.s.l. that is [0.5 - 2.5] km a.g.l. 
The discrepancy observations/simulations of around 1-2~m$\cdot$s$^{-1}$ (and a RMSE that can achieve 4~m$\cdot$s$^{-1}$)
is weak but, differently from the other discrepancies, 
seems to be the only discrepancy that does not come from the initial conditions. Where does it come from ? 
We did not observe any clear correlation between the discrepancy and the absolute value of the wind speed. 
We tested the sensitivity to the grid-point selection to check if a not precise selection of the grid-point of the summit could create some anomalous 
effects on the wind in the low atmosphere but we could exclude this hypothesis.
We note, however, that the radio-sounding is an in-situ measurement and that the balloon moves horizontally along the (x,y) plan during the ascension in the atmosphere. 
During the ascension time, the balloon therefore senses a volume of atmosphere shifted with respect to the zenithal direction. 
The radio-sounding lasts for around 4 minutes with a V$_{z}$~=~6~m$\cdot$s$^{-1}$ to reach the altitude of 4~km a.s.l. (the centre of the [3 - 5] km a.s.l. slab).
In this temporal interval the balloon can move somewhere (depending on the wind direction) in the (x,y) plane within a circle with a radius of $\sim$2.4~km. 
If we calculate the maximum variation of the wind speed ($\Delta$V) inside such a circle we see that, at 4~km a.s.l., $\Delta$V is of the 
order of 1.5-2 m$\cdot$s$^{-1}$. This is exactly the order of magnitude of the bias observed between simulations and observations.
Table~\ref{tab:3_5_wind_speed} reports these values for a few flights we analyzed. 

\begin{table}
\caption{\label{tab:3_5_wind_speed} Maximum variability ($\Delta$V) of the wind speed calculated at 4~km a.s.l. inside a circle in the plan (x,y) having a radius proportional to the wind speed observed at 4~km a.s.l. times 4 minutes (the time required to the ballon to reach 4~km a.s.l.).}
\begin{center}
\begin{tabular}{|c|c|c|c|}
\hline
Date & Hour (UT) & V$_{4km}$ (m$\cdot$s$^{-1}$)& $\Delta$V  (m$\cdot$s$^{-1}$)\\
\hline
1/8/2009 & 00:00 & 5  & 1.6 \\
11/11/2009 & 12:00 & 10 & 2 \\
19/11/2009 & 06:00& 5 & 1.4 \\
19/11/2009 & 12:00 & 10 & 1.6 \\
14/11/2009 & 12:00& 5 & 2 \\
\hline
\end{tabular}
\end{center}
\end{table}

The spatial horizontal inhomogeneity decreases with the height and disappears in the high part of the atmosphere. 
In other words, being that the horizontal distribution in the (x,y) plane of the wind speed in the low part of the atmosphere is not necessarily homogeneous, particularly above mountain regions, this could explain the discrepancy with the simulations in the [3~km - 5~km] a.s.l. range.
This argument tells us that the radio-sounding is not an optimal reference for comparisons with simulations to be done in the low part of the atmosphere.  
A preferable choice should be an instrument based on an optical remote sensing principle. 
To support this thesis we remind that in a recent study \citep{hagelin10}, a similar comparison of simulated versus measured wind speed 
obtained with a remote sensing instrument (a Generalized SCIDAR used for the wind speed measurements) provided a 
difference between the average wind speed vertical profiles better than 1 m$\cdot$s$^{-1}$ in the [0.5 - 1] km a.g.l. vertical range.

\section{Conclusions}
\label{conc}

In this paper we present the overview of the extended feasibility study (MOSE project) aiming at evaluating the Meso-Nh model ability in reconstructing all the main classical atmospherical parameters (temperature, wind speed, wind direction, relative humidity) as well as the optical turbulence ($\CN2$ and main integrated astro-climatic parameters: seeing, isoplanatic angle, wavefront coherence time). This is the first paper of a series that aims at summarizing the main results obtained in the context of this project. In this paper we focused our attention on the ability of the model in reconstructing the vertical stratification of the atmospheric parameters from a statistical point of view as well as analyzing model performances night by night on a sample of 50 radiosoundings distributed on 23 nights (11 nights in summer and 12 nights winter time). 
We proved that the Meso-Nh mesoscale model, using a configuration made of three imbricated domains (horizontal resolution of 10~km, 2.5~km and 0.5~km), provides a vertical distribution of wind speed and direction, temperature and relative humidity in the vertical slab [3 - 20] km a.s.l. with satisfactory levels of correlation with observations. \\
The wind direction has a bias within a few degrees and a RMSE within 10$^{\circ}$ between 5~km and 18~km a.s.l. Below 5~km and above 18~km the bias reaches 20$^{\circ}$ and the RMSE 60$^{\circ}$. \\
The wind speed in the [5-15]~km slab has a bias within 1 m$\cdot$s$^{-1}$ and a RMSE within 3 m$\cdot$s$^{-1}$. Above 15~km and in the [3 - 5] km a.s.l. range the bias is witihn 2~m$\cdot$s$^{-1}$. \\
The potential temperature has a bias within $\sim$ 2$^{\circ}$C from the ground up to 13~km a.s.l. Above 13~km the bias is within 4$^{\circ}$C. \\
The relative humidity has a bias within 10$\%$ all along the 20~km. Similar value is found for the RMSE with exception of the range [10 -14]~km a.s.l. in which the RMSE is within 15$\%$.
We showed that the deterioration of the model in the [3 -5]~km slab for the wind speed (bias passes from 1~m$\cdot$s$^{-1}$ to 2~m$\cdot$s$^{-1}$) is not necessarily due to the model itself but, highly probably, to the fact that our reference (balloon) is an in-situ measurement (see discussion in Section \ref{indiv}). We provided evidences that basically all the residual discrepancies between observations and model (in the high part of the atmosphere, typically h $>$ 5~km) are not due to intrinsic characteristics of the model itself (numerical schemes, physical packages, etc.) and they are therefore highly probably due to the initial conditions. Considering the good model reliability in reconstructing the wind speed vertical stratification the Meso-Nh model appears to be as the most practical and cheap tool, at present, to estimate and predict the temporal evolution of the wind speed all along the night. This method offers an advantageous solution in terms of accuracy and temporal coverage with respect to the General Circulation Models that are not reliable in the low part of the atmosphere and that provide information only at synoptic hours (00:00, 06:00, 12:00 and 18:00) UT.  It also provides considerable advantages with respect to others instruments (Generalized SCIDAR and/or Stereo SCIDAR). They are indeed both instruments that require a telescope of at least 1 m size and can hardly be used as automatic monitors. 

\section*{Acknowledgements}
This study is co-funded by the ESO contract: E-SOW-ESO-245-0933 (MOSE Project).
Meteorological data-set from the Automatic Weather Station (AWS) and mast at Cerro Armazones are from the Thirty Meter Telescope Site Testing -
Public Database Server \citep{schoeck09}.
Meteorological data-set from the AWS and mast at Cerro Paranal are from ESO Astronomical Site Monitor (ASM - Doc.N. VLT-MAN-ESO-17440-1773). 
We are very grateful to the whole staff of the TMT Site Testing Working Group for providing information about their data-set as well as 
to the ESO Board of MOSE (Marc Sarazin, Pierre-Yves Madec, Florian Kerber and Harald Kuntschner) for their constant support to this study. We acknowledge M. Sarazin and F. Kerber for providing us the ESO data-set used in this study.  
A great part of simulations are run on the HPCF cluster of the European Centre for Medium Weather Forecasts (ECMWF) - Project SPITFOT.

\appendix
\section{Radiosoundings sample}
\label{annex_a}

Table \ref{tab:annexe_23n} reports the list of radiosoundings launched at Cerro Paranal in two different periods of the year. In several cases, more than one radiosoundings has been launched during the same night.

\begin{table*}
 \centering
 \begin{tabular}{|c|c|c|c|c|}
 \hline
 \multicolumn{5}{|c|}{Simulated Nights - radiosoundings observations} \\
 \hline
  {\small 2009-07-29}         & {\small 2009-07-30  }       & {\small 2009-07-31}               & {\small 2009-08-01}                     & {\small 2009-08-02       }                                       \\
  {\scriptsize 12:00 (UT)} & {\scriptsize 12:15 (UT)} & {\scriptsize 00:30 12:00 (UT)} & {\scriptsize 00:00 06:15 12:00 (UT)} & {\scriptsize 00:00 06:15 (UT)}                               \\
 \hline
  {\small 2009-08-04 }        & {\small 2009-08-05 }              & {\small 2009-08-06}               & {\small 2009-08-07}                     & {\small 2009-08-08}                                        \\
  {\scriptsize 00:00 (UT)} & {\scriptsize 01:00 06:00 (UT)} & {\scriptsize 01:00 06:15 (UT)} & {\scriptsize 00:00 06:20 12:00 (UT)} & {\scriptsize 00:00 06:00 (UT)}                         \\
 \hline
  {\small 2009-08-09  }       & {\small 2009-08-10}               & {\small 2009-11-09 }        & {\small 2009-11-10       }  & {\small 2009-11-11 }                                                         \\
  {\scriptsize 00:00 (UT)} & {\scriptsize 01:10 12:00 (UT)} & {\scriptsize 12:00 (UT)} & {\scriptsize 12:00 (UT)} & {\scriptsize 00:00 06:00 12:00 (UT)}                                     \\
 \hline
  {\small 2009-11-12   }                  & {\small 2009-11-13}                     & {\small 2009-11-14 }                    & {\small 2009-11-15}                     & {\small 2009-11-16           }     \\
  {\scriptsize 00:00 06:00 12:15 (UT)} & {\scriptsize 00:00 06:00 12:00 (UT)} & {\scriptsize 00:00 06:00 12:00 (UT)} & {\scriptsize 00:00 06:00 12:20 (UT)} & {\scriptsize 00:00 06:15 (UT)} \\
 \hline
  {\small 2009-11-17    }                 & {\small 2009-11-18}                     & {\small 2009-11-19  }                   &  &  \\
  {\scriptsize 00:00 06:00 12:00 (UT)} & {\scriptsize 00:00 06:00 12:00 (UT)} & {\scriptsize 00:00 06:00 12:00 (UT)} &  &  \\ 
 \hline
 \end{tabular}
 \caption{List of the 23 simulated nights with available radiosoundings observations in 2009. Below each date are reported the hours at which a balloon was launched. 
  \label{tab:annexe_23n}}
\end{table*}

\clearpage

\section{Wind speed vertical profiles: model vs. observations}
\label{annex_b}

\begin{figure*}
\begin{center}
\begin{tabular}{c}
\includegraphics[width=0.65\textwidth]{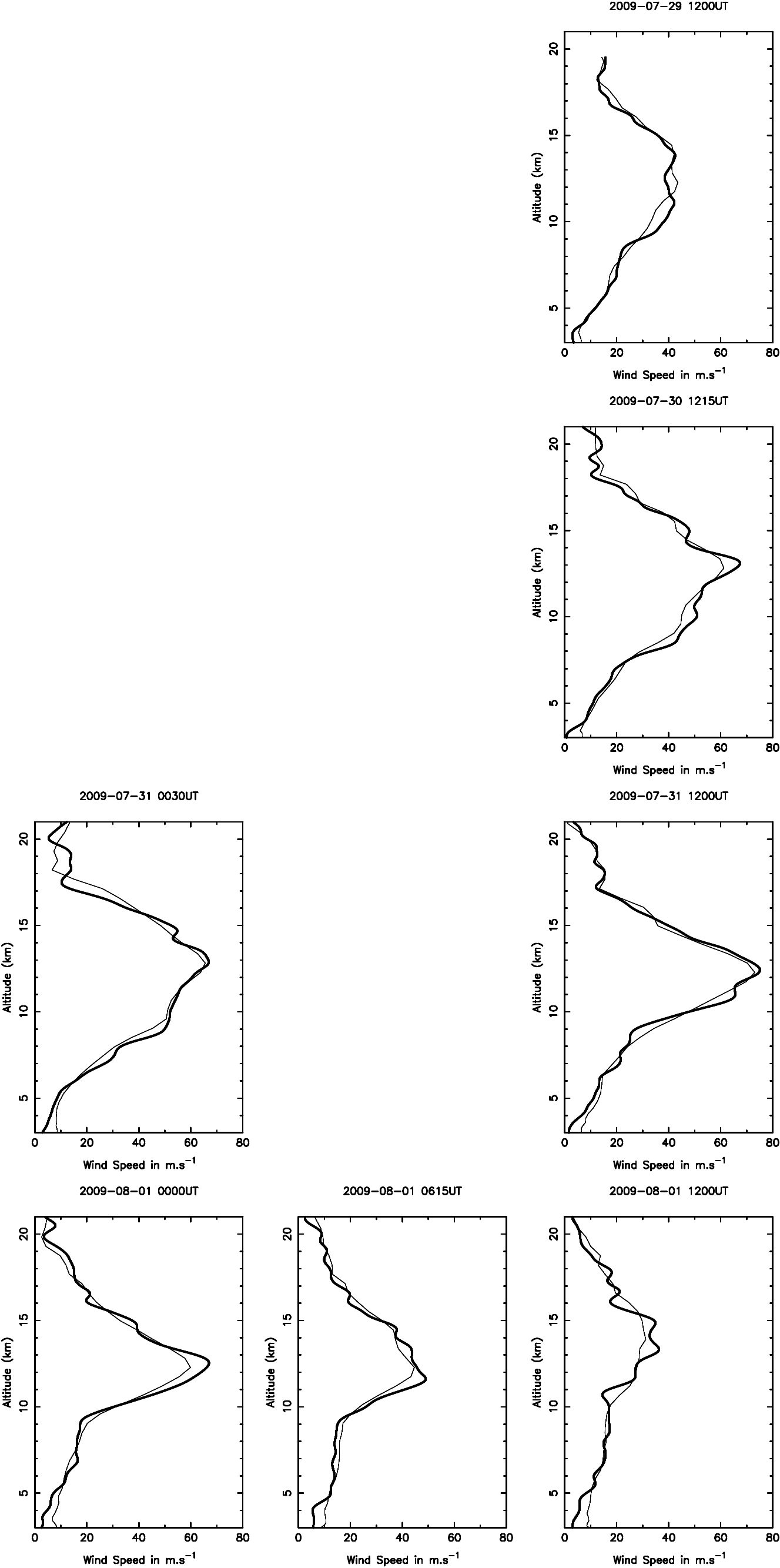}
\end{tabular}
\end{center}
\caption{Wind speed vertical profiles observed from radiosondes measurements (thick line) and simulated by the model (thin line). Each row is a different night, and 
all the nights from the Table~\ref{tab:annexe_23n} are represented. Unit in m$\cdot$s$^{-1}$.} 
\label{fig:annex_1} 
\end{figure*}

\begin{figure*}
\begin{center}
\begin{tabular}{c}
\includegraphics[width=0.435\textwidth]{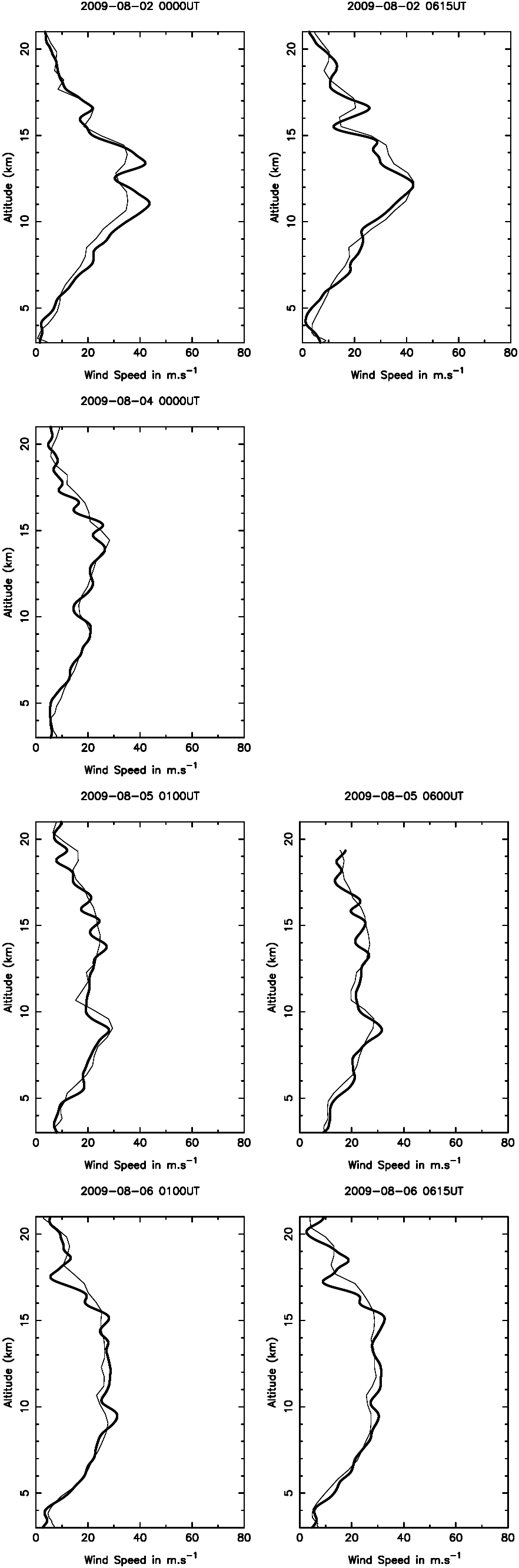}
\end{tabular}
\end{center}
\caption{Following of Fig.~\ref{fig:annex_1}.}
\label{fig:annex_2} 
\end{figure*}

\begin{figure*}
\begin{center}
\begin{tabular}{c}
\includegraphics[width=0.65\textwidth]{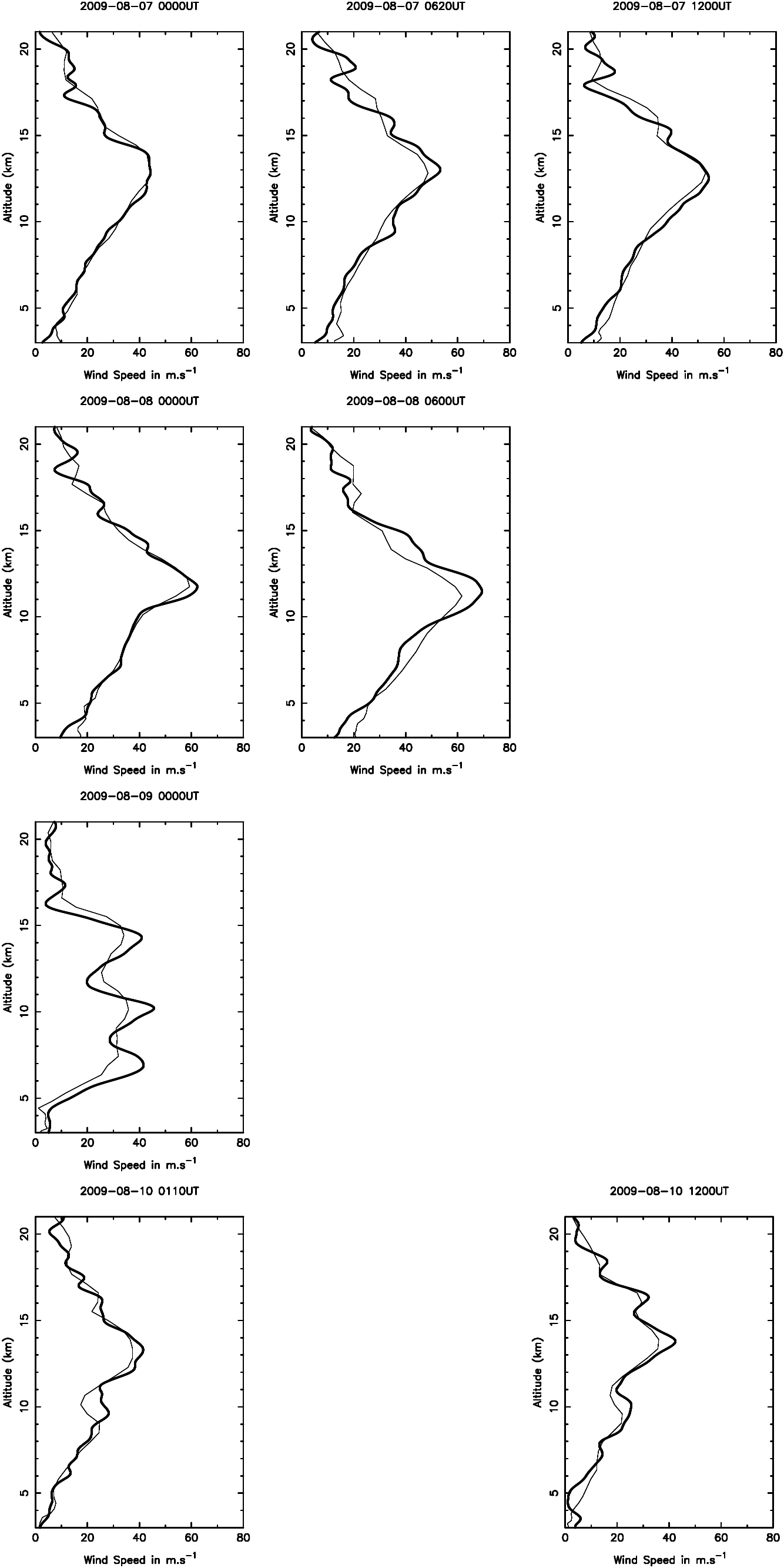}
\end{tabular}
\end{center}
\caption{Following of Fig.~\ref{fig:annex_1}.}
\label{fig:annex_3} 
\end{figure*}

\begin{figure*}
\begin{center}
\begin{tabular}{c}
\includegraphics[width=0.65\textwidth]{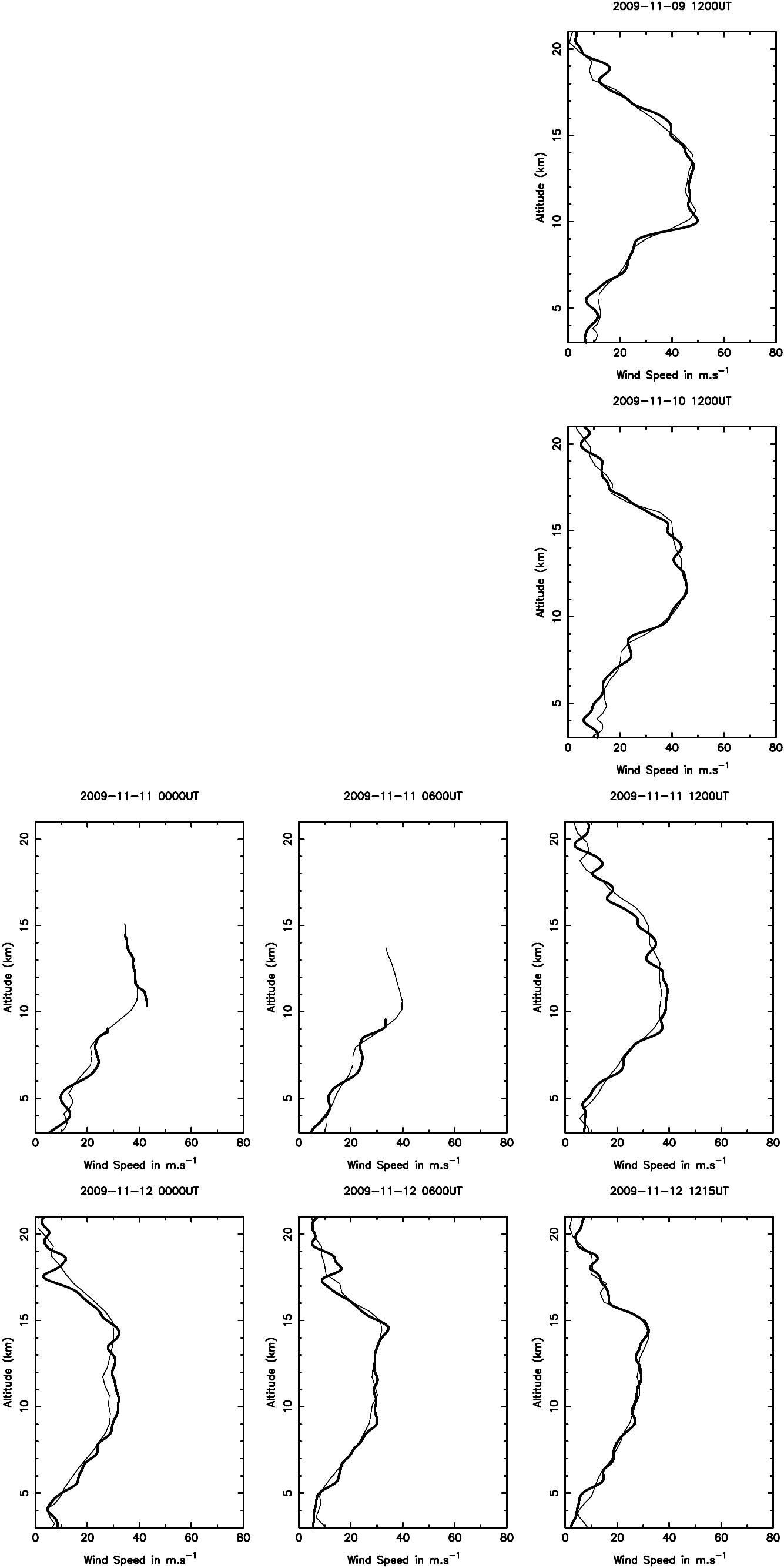}
\end{tabular}
\end{center}
\caption{Following of Fig.~\ref{fig:annex_1}.}
\label{fig:annex_4} 
\end{figure*}

\begin{figure*}
\begin{center}
\begin{tabular}{c}
\includegraphics[width=0.65\textwidth]{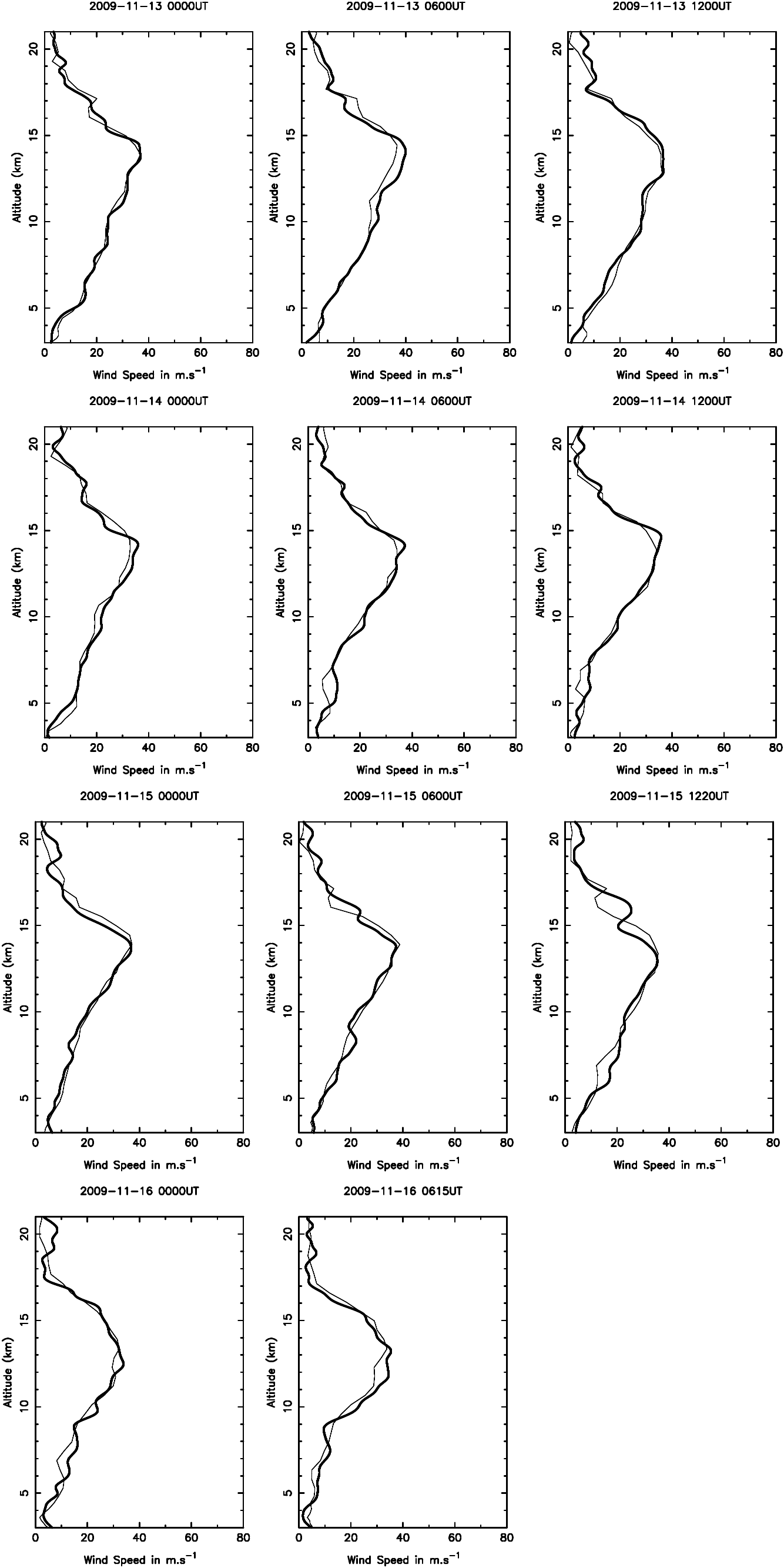}
\end{tabular}
\end{center}
\caption{Following of Fig.~\ref{fig:annex_1}.}
\label{fig:annex_5} 
\end{figure*}

\begin{figure*}
\begin{center}
\begin{tabular}{c}
\includegraphics[width=0.65\textwidth]{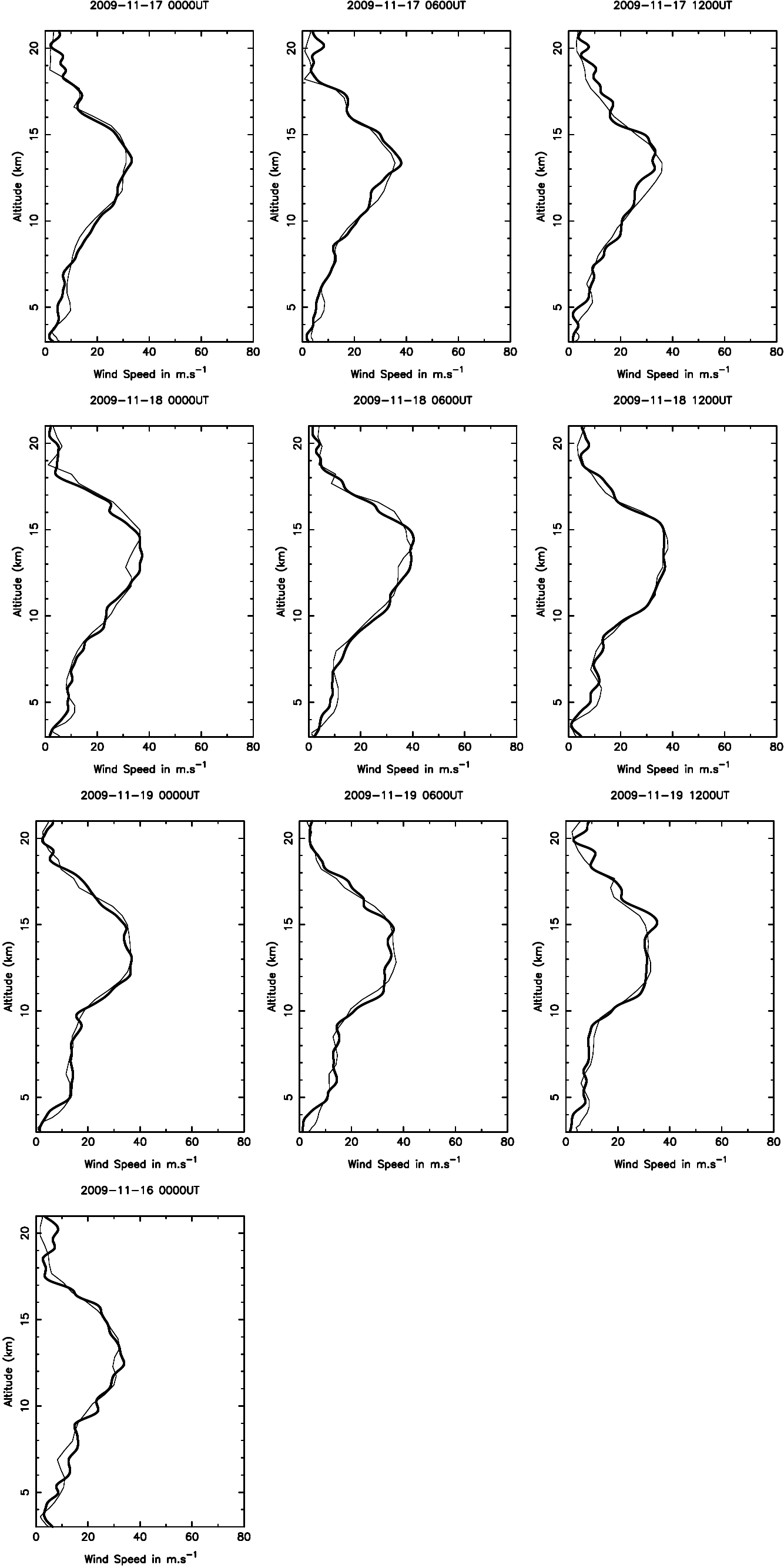}
\end{tabular}
\end{center}
\caption{Following of Fig.~\ref{fig:annex_1}.}
\label{fig:annex_6} 
\end{figure*}

Fig.\ref{fig:annex_1}-Fig.\ref{fig:annex_6} show the 50 different vertical profiles of the wind speed as measured by radiosoundings (thick lines) and simulated by the model (thin lines). Each row is a different night of the total sample of the total 23 nights (see Section \ref{gen}). In this set of figures is evident how the model well reconstructs the wind speed features at all heights above the ground and in most of the nights and the individual interval of time corresponding to the balloon launch. 

\newpage

\newpage

\newpage

\newpage

\newpage
%%%%%%%%%%%%%%%%%%%%%%%%%%%%%%%%%%%%%%%%%%%%%%%%%%%%%%%%%%%%%%%%%%%

\label{lastpage}
\end{document}